\documentclass[twocolumn,twoside,slac_two]{revtex4}
\usepackage{graphicx}
\usepackage{fancyhdr}
\begin{document}

\title{Galaxy Clusters and Dark Matter Properties}

%

\author{J. S. Arabadjis, M. W. Bautz}
\affiliation{MIT, Cambridge, MA 02139, USA}

\begin{abstract}

Laboratory experiments, large-scale computer simulations and observational
cosmology have begun to make progress in the campaign to identify the particle
responsible for gravitationally-inferred dark matter.  In this contribution we
discuss the dark matter density profiles in the cores of  nearby galaxy
clusters and estimate the gamma-ray flux expected for MSSM dark matter over a
range of neutralino masses.

\end{abstract}

\maketitle

\thispagestyle{fancy}


\section{INTRODUCTION}

Galaxy clusters are dark matter-dominated objects.  Since they are believed to
constitute a nearly fair sample of the matter content of the universe
\cite{white}, WMAP results suggest that they are comprised of roughly 15\%
baryons and 85\% non-baryonic dark matter \cite{wmap-spergel}.  Numerical
experiments which simulate their formation via the hierarchical assembly of
cold dark matter (CDM) halos suggest that their density profiles are adequately
described by a pair of power laws and a transition radius
\cite{nfwsix,nfwseven,moore}.  Indeed, the dark matter density profile in the
centers of relaxed clusters shows power-law behavior to scales as small as
$\sim10$ kpc \cite{abg,bl,ab}.  If the density profile remains a power law to
small enough radius, the central density may become large enough that dark
matter self-annihilation produces a gamma-ray flux observable with current
instrumentation.

The neutralino is perhaps the leading candidate for the dark matter particle
\cite{jkg}.  The observational signal for neutralino annihilations can be
quite spectacular.  Two of the annihilation channels result in the production
of monochromatic gamma rays, unlikely to be confused with other astrophysical
processes.  Other channels can lead to substantially more gamma rays, although
since they produce a continuous spectrum they are more difficult to distinguish
from high energy processes such as shock heating.

Most searches for the gamma ray signature of dark matter annihilation have
centered on local dark matter concentrations such as the Galactic center
\cite{gs} and halo \cite{bub,abo} and Local Group dwarf spheroidals \cite{efs}.
However, if the centers of galaxy clusters are cuspy to sufficienty small
scale, they too may be observable with the current generation of gamma ray
telescopes.  In this contribution we calculate the expected annihilation
signal from clusters at distances less than $\sim 100$ Mpc for core density
profiles determined through X-ray observations.

\section{NEUTRALINO ANNIHILATION}

We will assume throughout this contribution that the lightest supersymmetric
particle is the neutralino, a linear superposition of the superpartners of the
photon, Z$^0$, and neutral Higgs bosons. and that it provides the bulk of the
dark matter in the universe \cite{ehnos,jkg}.  The neutralino, a Majorana
fermion, self-annihilates in the early universe until the annihilation rate is
exceeded by the Hubble rate.  The neutralino relic density $\Omega_\chi$
depends upon its exact composition and the presence of resonances, but it can
be determined approximately using \cite{jkg}

\begin{equation}\label{eq:relic}
\Omega_\chi h^2 \simeq
\frac{3\times 10^{-27} \, {\rm cm}^3 \, {\rm s}^{-1}}{\langle\sigma v\rangle}
\end{equation}

\noindent where $h$ is the Hubble parameter and $\langle\sigma v\rangle$ is the
thermally averaged annihilation cross section.  Assuming neutralinos are by far
the cosmologically dominant dark matter species, then
$\Omega_\chi \simeq (1-f_b) \, \Omega_m$, where $f_b$ is the matter baryon
fraction and $\Omega_m$ is the total matter density.  For $h=0.67$ and
$f_b=0.15$, the total neutralino annihilation cross section is
$\langle\sigma v\rangle = 3\times 10^{-26}$ cm$^3$ s$^{-1}$.

Neutralinos can annihilate directly into a pair of monochromatic gamma rays
[$\gamma\gamma$], a gamma and a neutral Z boson [$\gamma$Z$^0$], or into a
gamma continuum through a plethora of hadronization processes [$\gamma$(h)].
The dominant ``channel'' is hadronization, with a resulting gamma ray spectrum
conveniently approximated by \cite{abo}:

\begin{equation}\label{eq:gh}
\frac{dN_\gamma}{dE_\gamma} = \frac{5}{4 m_\chi} \int_{x_{min}}^{1} dx \,
\frac{(1-x)^2}{x^{3/2}(x^2-\eta^2)^{1/2}}
\end{equation}

\begin{figure}
\includegraphics[width=65mm]{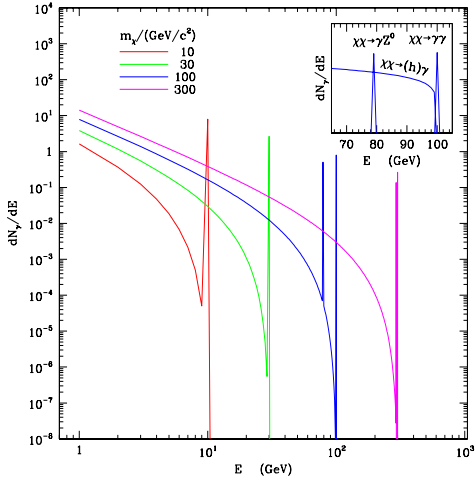}
\caption{Gamma ray annihilation spectra for four different neutralino masses.
The spectra shown are, left to right, for 10, 30, 100 and 300 GeV/c$^2$.  The
inset shows the individual contributions from the $\gamma\gamma$, $\gamma$Z$^0$
and $\gamma$(h) processes.
\label{fig:spectrum}}
\end{figure}

\noindent Here $x = E_\pi/m_\chi$, $\eta = m_\pi/m_\chi$, and
$x_{min} = E_\gamma + m_\chi \eta^2/4E_\gamma$.  The neutral pion and
neutralino masses are $m_\pi$ and $m_\chi$.  The $\gamma\gamma$ and
$\gamma$Z$^0$ spectra are given by

\begin{equation}\label{eq:gg}
\frac{dN_\gamma}{dE_\gamma} = \frac{2}{E} \,\,
\delta\left(1-\frac{m_\chi}{E_\gamma}\right)
\end{equation}

\noindent and

\begin{equation}\label{eq:gz}
\frac{dN_\gamma}{dE_\gamma} = \frac{1}{E} \,\,
\delta\left(1-\frac{m_\chi(1-(m_{Z^0}/2m_\chi)^2)}{E_\gamma}\right)
\end{equation}

\noindent respectively \cite{bu,ub,kzw}.  Estimates for the cross section of
these last two channels vary over several orders of magnitude.  We use the
following values for our analysis:
$\langle\sigma v\rangle_{\gamma({\rm h})} = 3\times 10^{-26}$ cm$^3$ s$^{-1}$,
$\langle\sigma v\rangle_{\gamma\gamma} = 
 \langle\sigma v\rangle_{\gamma {\rm Z}^0} = 0.01
 \langle\sigma v\rangle_{\gamma({\rm h})}$.  These spectra are shown in
Figure~\ref{fig:spectrum} for four
values of the neutralino mass.

\section{GAMMA RAY LUMINOSITY OF CLUSTERS}

The annihilation radiation luminosity of a galaxy cluster is given by

\begin{equation}\label{eq:lum}
L_\gamma = \frac{N_\gamma \langle\sigma v\rangle}{4\pi m_\chi^2}
\int \, \rho^2 \, dV
\end{equation}

\noindent where $\rho$ is the matter density of neutralinos, and the integral
is taken over the entire cluster volume.  For a dark matter core profile
following $\rho = k r^\alpha$, this integral diverges toward the center for
$\alpha \le -1.5$.  Although astrophysical processes will probably impose a
density cut-off, significant gamma ray signatures may still result.

\begin{figure}
\includegraphics[width=65mm]{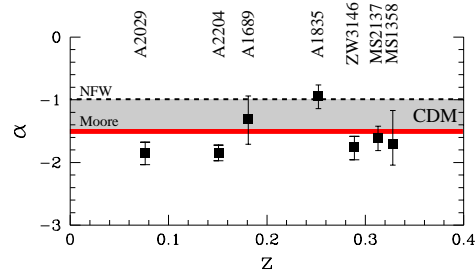}
\caption{Inner logarithmic density slopes for seven relaxed galaxy clusters
\cite{aba,ab}.  The gray band, bracketed by the NFW and Moore profiles,
represents the prediction of standard CDM cosmology.
\label{fig:alpha}}
\end{figure}

Figure~\ref{fig:alpha} shows the inner logarithmic density slope for a sample
of seven relaxed clusters observed with the Chandra X-ray Observatory
\cite{aba,ab}.  The slopes generally span the range $-1\le \alpha\le -2$,
suggesting that gamma ray luminosities may indeed be quite high at the center
of some clusters.

\begin{figure*}[t]
\centering
\includegraphics[width=135mm]{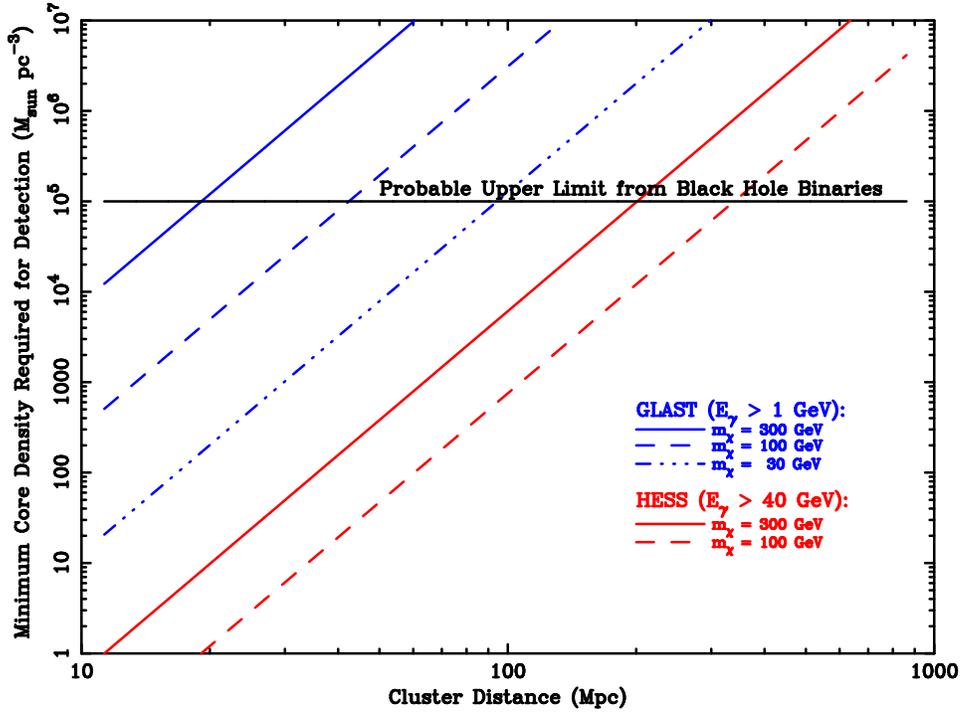}
\caption{Dark matter central density required for detectability as a function
of cluster distance.  The upper three diagonal lines represent GLAST
sensitivity limits for $m_\chi$ = 300, 100, 30 GeV, the lower two for HESS
sensitivity limits at 300 and 100 GeV.
\label{fig:rhomin_detect}}
\end{figure*}

The largest astrophysically possible neutralino density is determined by the
free-fall timescale at the center, with the density determined by equating the
free-fall rate

\begin{equation}\label{eq:tf}
t_{f}^{-1} = \sqrt{32G\rho/3\pi}
\end{equation}

\noindent to the annihilation rate

\begin{equation}\label{eq:ta}
t_a^{-1} = n_\chi \langle\sigma v\rangle
\end{equation}

\noindent since neutralinos cannot annihilate faster than they are supplied to
the high density regions.  This implies a free-fall-limited maximum density

\begin{equation}\label{eq:pf}
\rho_{max,f} = 1.2\times 10^{21} \left( \frac{m_\chi}{{\rm 10 \, GeV/c}^2}
\right)^2 \, \, {\rm M}_\odot \, {\rm pc}^{-3}
\end{equation}

\noindent However, dynamical heating of neutralinos by the formation of
supermassive black hole binaries may reduce this value significantly, to about
$10^5$ M$_\odot$ pc$^-3$ \cite{merrittetal}.  (Other processes such as
gravitational heating \cite{merritt} may be less restrictive.)
Figure~\ref{fig:rhomin_detect} illustrates the peak central density required
for a galaxy cluster to be rendered detectable by GLAST and HESS as a function
of cluster distance.  The cluster central density profile used is $\alpha=-1.5$
to a cut-off radius, where it remains at the limiting density.  The horizontal
black line indicates a crude estimate of the upper limit of the density imposed
by the formation of supermassive black holes during the growth of the central
cluster galaxy \cite{merrittetal}.  This figure shows that {\it if central
densities reach high, but plausible values, then annihilation radiation from
neutralino dark matter may be detectable to distances of tens to hundreds of
Mpc.}  This radiation would carry important information about the very centers
of these objects.

\section{A TALE OF TWO CLUSTERS}

\begin{figure*}[t]
\centering
\includegraphics[width=135mm]{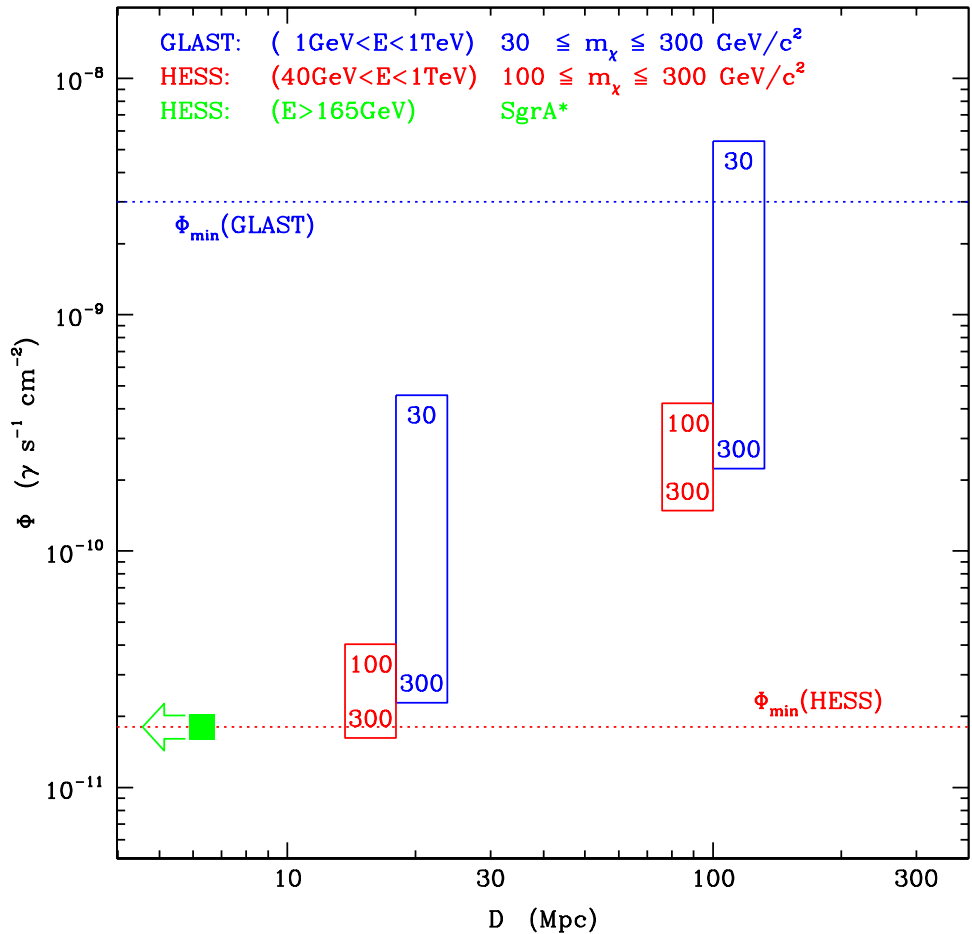}
\caption{Expected gamma ray fluxes from Virgo (left at 17 Mpc) and a more
distant, but steeper-profiled, cluster at 100 Mpc.  For reference the GLAST
and HESS detection thresholds are shown, as is the recent detection of Sgr A*
\cite{aharonian}.
\label{fig:fluxes}}
\end{figure*}

It makes sense, in light of these results, to ask the question:  Could any
{\it real} clusters be detected in their annihilation radiation?  To address
this question we use two examples.  The first is the Virgo Cluster at a
distance of 17 Mpc, with a central density slope of $-1.3$ \cite{koop}.  For
the second example we use an Abell 2029-like cluster ($\alpha=-2$; \cite{ab})
at a distance of 100 Mpc.  The annihilation radiation flux from a cluster, in
photons s$^{-1}$ cm$^{-2}$ sr$^{-1}$, is

\begin{eqnarray}\label{eq:flux}
F_\gamma & = & 2.9\times 10^{-7} \, \left(
\frac{\langle\sigma v\rangle}{3\times 10^{-26} \, {\rm cm}^3 \, {\rm s}^{-1}}
\right) \nonumber \\
& & \cdot \, \left( \frac{m_\chi}{10 {\rm GeV/c}^2} \right)^{-2}
\frac{1}{\Omega} \, \int_{\Omega} d\Omega \, \int_{E} N_\gamma \nonumber \\
& & \cdot \, \int_{\rm LOS} \left( \frac{\rho}{10^{-24} {\rm g/cm}^3} \right)^2
\frac{dl}{{\rm 1 \, kpc}}
\end{eqnarray}

\noindent where $\Omega$ is the beam size.  Figure~\ref{fig:fluxes} shows the
expected annihilation radiation flux from each object as seen by GLAST (for
$30 \le m_\chi \le 300$ GeV/c$^2$) and HESS (for $100 \le m_\chi \le 300$).
The GLAST and HESS detection thresholds are shown for comparison, as is the
tentative HESS detection of Sgr A* \cite{aharonian}.  In both cases we have
adopted $\rho_{max} = 10^5$ M$_\odot$ pc$^{-3}$.  The fluxes have been
calculated from the surface brightness integral over a beam size of $10^{-5}$
sr.  GLAST may detect an annihilation signal from a steep-core cluster if the
neutralino mass is low; HESS may be able to detect it for a wider range of
distance and particle mass.

\section{CONCLUSION}
We have shown that, for plausible values of the central dark matter slope and
for sensible astrophysical constraints on the peak density, neutralino
annihilation radiation from the centers of galaxy clusters may be detectable
using current and near-future gamma ray telescopes.  These signals would not
only provide spectacular, if indirect, evidence for the existence of dark
matter, but they would also carry information about conditions at the centers
of galaxy clusters.

\bigskip 
\begin{acknowledgments}
The authors wish to thank Sergio Colafrancesco for his valuable perspective.
This work was supported by NASA grant 2834-MIT-SAO-4018.
\end{acknowledgments}

\bigskip 

\end{document}